\documentclass[12pt, reqno]{amsart}

\usepackage[margin=.9in, a4paper, foot=.3in]{geometry}

\let\oldtocsection=\tocsection

\let\oldtocsubsection=\tocsubsection

\let\oldtocsubsubsection=\tocsubsubsection

\renewcommand{\tocsection}[2]{\hspace{0em}\oldtocsection{#1}{#2}}
\renewcommand{\tocsubsection}[2]{\hspace{1em}\oldtocsubsection{#1}{#2}}
\renewcommand{\tocsubsubsection}[2]{\hspace{2em}\oldtocsubsubsection{#1}{#2}}

\usepackage{cite}

\usepackage{graphicx, color}

\definecolor{darkblue}{rgb}{0,0,.6}
\definecolor{darkred}{rgb}{.6,0,0}
\definecolor{darkgreen}{rgb}{0,0.6,0}

\usepackage{hyperref}
\hypersetup {
    pdfstartview=FitH,
    colorlinks,
    citecolor=darkgreen,
    linkcolor=darkred,
    urlcolor=darkblue
}

\usepackage[noBBpl,slantedGreek]{mathpazo}
\usepackage[mathcal]{euscript}

\usepackage{amssymb}

\makeatletter
\def\section{\@startsection{section}{1}%
  \z@{-.7\linespacing\@plus -\linespacing}{.5\linespacing}%
  {\normalfont\scshape\centering}}
\def\subsection{\@startsection{subsection}{2}%
  \z@{-.5\linespacing\@plus -.7\linespacing}{.5em}%
  {\normalfont\bfseries\mathversion{bold}}}
\makeatother

\allowdisplaybreaks

\numberwithin{equation}{section}

\newcommand {\bbC}{\mathbb C}

\newcommand {\bbZ}{\mathbb Z}

\newcommand {\calA}{\mathcal A}
\newcommand {\calB}{\mathcal B}
\newcommand {\calC}{\mathcal C}
\newcommand {\calD}{\mathcal D}
\newcommand {\calE}{\mathcal E}

\newcommand {\calL}{\mathcal L}
\newcommand {\calM}{\mathcal M}
\newcommand {\calQ}{\mathcal Q}
\newcommand {\calR}{\mathcal R}
\newcommand {\calT}{\mathcal T}

\newcommand {\gothb}{\mathfrak b}
\newcommand {\gothh}{\mathfrak h}
\newcommand {\gothg}{\mathfrak g}
\newcommand {\gothk}{\mathfrak k}

\newcommand {\gothgl}{\mathfrak{gl}}

\newcommand {\gothsl}{\mathfrak{sl}}

\def \qbinom #1#2{\begin{bmatrix} #1 \\ #2 \end{bmatrix}}
\newcommand {\mbar}[3]{\hskip #2 \overline{\hskip -#2 #1 \hskip -#3} \hskip #3}

\newcommand {\ovW}{\mbar{W}{.03em}{.03em}}
\newcommand {\ovrho}{\mbar{\rho}{.03em}{.03em}}
\newcommand {\ovvarphi}{\mbar{\varphi}{.03em}{.03em}}

\newcommand {\ovcalT}{\mbar{\mathcal T}{.03em}{.03em}}
\newcommand {\ovcalQ}{\mbar{\mathcal Q}{.03em}{.03em}}

\newcommand {\Osc}{\mathrm{Osc}}

\newcommand {\sgn}{\mathrm{sgn}}

\newcommand {\uqbp}{\mathrm U_q(\mathfrak b_+)}

\DeclareMathOperator {\End}{End}
\DeclareMathOperator {\id}{\mathrm{id}}

\DeclareMathOperator {\tr}{tr}

\title{Quantum affine algebras and universal functional relations}

\author[Kh. S. Nirov]{\vskip .2em Khazret S. Nirov}
\address{Institute for Nuclear Research of the Russian Academy of Sciences, 60th October Ave 7a, 117312 Moscow, Russia}
\curraddr{Fachbereich C -- Physik, Bergische Universit\"at Wuppertal, 42097 Wuppertal, Germany}
\email{nirov@uni-wuppertal.de}

\author[A. V. Razumov]{Alexander V. Razumov}
\address{Institute for High Energy Physics, NRC ``Kurchatov Institute'', 142281 Protvino, Moscow region, Russia}
\email{Alexander.Razumov@ihep.ru}

\begin{document}

\addtolength {\jot}{3pt}

\begin{abstract}
By the universal integrability objects we mean certain monodromy-type and transfer-type
operators, where the representation in the auxiliary space is properly fixed, while the 
representation in the quantum space is not. This notion is actually determined by the 
structure of the universal R-matrix. We call functional relations between such universal 
integrability objects, and so, being independent of the representation in the quantum space, 
the universal functional relations. We present a short review of the universal functional 
relations for the quantum integrable systems associated with the quantum groups of loop Lie 
algebras. 
\end{abstract}

\maketitle

\tableofcontents

\section{Introduction} \label{s:0}

The discovery of the so-called {\em Hidden Grassmann Structure\/} in the state space of the $XXZ$ model \cite{BooJimMiwSmiTak07, BooJimMiwSmiTak09, JimMiwSmi09, BooJimMiwSmi10, JimMiwSmi11} opened principally novel perspectives in the study of correlation functions in quantum integrable systems. At the same time, it raised fundamental questions about the algebraic origin of the fermionic basis in the $XXZ$ model and its possible extensions. 
Trying to give any adequate response to this challenge, we came to the conclusion that 
we need a better understanding of Baxter's $TQ$-relations \cite{Bax82, Bax04}, with all 
their ingredients and possible generalizations. Then, at certain point of our study, we 
came across an earlier series of remarkable papers on integrable structure of conformal 
field theory \cite{BazLukZam96, BazLukZam97, BazLukZam99, BazHibKho02}. 
For us, the main idea of these pioneer works was that the objects describing the model and related to its integrability should be constructed on the basis of the {\em universal $R$-matrix\/}, which is an element of the tensor product of two copies of the quantum group \cite{ChaPre91}. Traditionally, the first factor in this tensor product is regarded as the {\em auxiliary space\/}, and the second one as the {\em quantum space\/}. But their roles can certainly be interchanged, giving rise to various interesting relations between different integrable structures. The functional relations, supposed to be a substitute and generalization of the famous Bethe Ansatz \cite{Bet31}, should follow from the characteristics of the appropriate representations of the underlying quantum group. Altogether, our deliberations over the contents of these papers made clear that, in the theory of quantum integrable systems, the functional relations are too important to treat them superficially. And so, we have carefully reconsidered the whole quantum group approach to quantum integrable systems. Essential part of this revision is actually what we are presenting here in a short review based mainly on our publications \cite{BooGoeKluNirRaz10, BooGoeKluNirRaz11, BooGoeKluNirRaz13, BooGoeKluNirRaz14a, BooGoeKluNirRaz14b, 
NirRaz14}.

By $\calL(\gothg)$ we denote the loop Lie algebra of a finite-dimensional Lie algebra 
$\gothg$ and by $\widetilde \calL(\gothg)$ its standard central extension \cite{Kac90}. 
The corresponding quantum group is denoted by adding the symbol $U_q$, as appropriate 
for the $q$-deformation of the universal enveloping algebra of a given Lie algebra. It is assumed here that $q$ is not a root of unity. To construct integrability objects one uses spectral parameters. They are introduced by means of a $\bbZ$-gradation of the quantum group under consideration. In the case of our interest, the quantum group $U_q(\calL(\gothsl_n))$, 
a $\bbZ$-gradation is determined by $n$ integers $s_i$, $i = 0,\ldots,n-1$, and we use the notation $s = s_0 + \ldots + s_{n-1}$ for their total sum. Hence, the functional relations 
are certain difference equations for the integrability objects considered as functions of the spectral parameters. When the Bethe Ansatz is not applicable anymore, the functional relations can serve to diagonalize the transfer matrix of such models. However, they can also be studied 
independently to elucidate certain algebraic structures of quantum integrable systems. In particular, they can help in understanding the group-theoretic background of the Hidden Grassmann Structure and see ways to extend it to other models. It turns out here that it 
is justified to fix the auxiliary space by an appropriate representation, while keeping the quantum space free. Possible advantage of such an approach was anticipated and mentioned earlier in different contexts, see, for example, \cite{AntFei97, BazTsu08}. In our recent 
works \cite{BooGoeKluNirRaz13, BooGoeKluNirRaz14a, BooGoeKluNirRaz14b, NirRaz14}, we have developed and elaborated this idea and have given the complete proof of the functional relations in the form independent of the representation of the quantum group in the quantum space. We have also described the specialization of the universal functional relations to 
the case when the quantum space is the state space of a discrete spin chain. In this
presentation we consider only the first part of the program, the universal integrability objects and the corresponding universal functional relations without their specialization 
in the quantum space.

We use the notation $\kappa_q = q - q^{-1}$, so that the definition of the $q$-number can be written as 
\begin{equation*}
[\nu]_q = \frac{q^\nu - q^{- \nu}}{q - q^{-1}} = \kappa_q^{-1}(q^\nu - q^{-\nu}), \qquad \nu \in \bbC.
\end{equation*}

\section{Quantum groups in general} \label{s:1}

Here we recollect basic definitions of a well-known object and fix the corresponding
notations. The object in question, denoted here by $\calA$, is nothing but the {\em 
quantum group\/}, being actually the main tool for the whole construction. We generally
follow the definitions proposed by Drinfeld \cite{Dri85, Dri87} and Jimbo \cite{Jim85};
for a review see \cite{ChaPre94}. Hence, it is a unital associative algebra obtained by 
a deformation of the universal enveloping algebra of a given Lie algebra. The quantum 
group is thus an algebra, and if the initial Lie algebra is an affine Kac--Moody Lie 
algebra, then one deals with a {\em quantum affine algebra\/}. This is the case of our 
consideration.

It should be noted here that, depending on the nature of the deformation parameter $q$, 
the quantum group can be interpreted differently. If $q = \exp \hbar$, where $\hbar$ is 
an indeterminate, then the quantum group is a $\bbC[[\hbar]]$-algebra. If $q$ itself is 
an indeterminate, then the quantum group is a $\bbC(q)$-algebra. If $q = \exp \hbar$,
where $\hbar$ is a complex number, then the quantum group is a $\bbC$-algebra. We deal
with the quantum group defined according to the last way in order to have meaningful
traces, see, for example, \cite{JimMiw95, EtiFreKir98}.

The quantum group is a Hopf algebra with respect to appropriate co-multiplication,
antipode and co-unit. Moreover, as any Hopf algebras, it also has an opposite 
co-\-mul\-tip\-li\-ca\-ti\-on $\Delta^{\mathrm{op}}$, defined by the initial 
co-multiplication $\Delta$ as 
\[
\Delta^{\mathrm{op}} = \Pi \circ \Delta, \qquad \Pi (a \otimes b) = b \otimes a,
\qquad a, b \in \calA.
\]
Now, assuming that there exists an invertible element $\calR$ of the tensor product 
of two copies of the quantum group, such that
\begin{equation}
\Delta^{\mathrm{op}}(a) = \calR \, \Delta(a) \calR^{-1}, \qquad 
\calR \in \calA \otimes \calA, \label{1.1}
\end{equation}
for any element $a$ of the quantum group, and the following relations
\begin{equation}
(\Delta \otimes \id) (\calR) = \calR^{13} \calR^{23}, \qquad 
(\id \otimes \Delta) (\calR) = \calR^{13} \calR^{12},   \label{1.2} 
\end{equation}
with the standard convention about the indices, are fulfilled, we restrict ourselves 
by the so-called {\em quasi-triangular\/} Hopf algebras.\footnote{In fact, there is 
a subtle conflict between the definition of the quantum group as a $\bbC$-algebra 
and the last requirement. However, this problem is resolvable in our case, see, for 
example, the discussions in \cite{Tan92, BooGoeKluNirRaz14a, BooGoeKluNirRaz14b, 
NirRaz14}.} The element $\calR$ is called the {\em universal $R$-matrix\/}. It 
follows from (\ref{1.1}) and (\ref{1.2}) that it satisfies the {\em master equation\/}
\begin{equation}
\calR^{12} \calR^{13} \calR^{23} = \calR^{23} \calR^{13} \calR^{12} 
\label{1.3}
\end{equation}
defined on the tensor product of three copies of the quantum group. However, it 
is important to note that in the case under consideration the universal $R$-matrix 
belongs actually to a smaller subset of the tensor square of the full quantum group; 
to be precise, to the tensor product of the Borel subalgebras $\calB_+$ and $\calB_-$ 
of the quantum group $\calA$, i.~e. $\calR \in \calB_+ \otimes \calB_- \subset \calA 
\otimes \calA$. Therefore, and in particular, the master equation is defined on the 
tensor product $\calB_+ \otimes \calA \otimes \calB_-$.

\section{Universal integrability objects} \label{s:2}

Let us introduce a group-like element, that is an invertible element of the quantum
group with the co-multiplication property given by
\[
\Delta(t) = t \otimes t, \qquad t \in \calA.
\]
With the help of the universal $R$-matrix we define the {\em universal integrability
objects\/}. These are the monodromy- and transfer-type operators defined as follows.

First, let $\varphi$ be a representation of the quantum group $\calA$ in a vector
space $V$\footnote{In other words, $V$ is an $\calA$-module.}
\[
\varphi : \calA \to \End(V).
\]
Then, the corresponding universal monodromy operator $\calM_\varphi$ is defined as 
\begin{equation}
\calM_\varphi = (\varphi \otimes \id) (\calR) \in \End(V) \otimes \calA.
\label{2.1}
\end{equation}
The corresponding universal transfer operator is defined as
\begin{equation}
\calT_\varphi = (\tr_V \otimes \id) (\calM_\varphi(\varphi(t) \otimes 1)) 
= ((\tr_V \circ \varphi) \otimes \id)(\calR (t \otimes 1)),
\label{2.2}
\end{equation}
where $1$ means the unit element of $\calA$, and now $t$ makes sense as a {\em twist element\/}.
Using appropriate $t$, in general, we can obtain meaningful traces over infinite-dimensional 
representation spaces, see, for example, \cite{BooGoeKluNirRaz14a}. The representation $\varphi$ can depend on certain parameters, which we collectively denote by $\zeta$. 
In such a case we write $\varphi_\zeta$ and $\calM_\varphi(\zeta)$, $\calT_\varphi(\zeta)$, respectively.  

The other universal integrability objects are introduced in a similar way, 
but the representation $\rho$ used for them is essentially different from 
the representation $\varphi$ used earlier. Indeed, here we recall that, in 
the case under consideration, the universal $R$-matrix is an element of the
tensor product of two principal Borel subalgebras of $\calA$, that is $\calR 
\in \calB_+ \otimes \calB_-$. Then, let $\rho$ be such a representation of
$\calB_+$ that it cannot be extended to a representation of the whole quantum
group $\calA$. Therefore, it cannot be obtained simply by the 
restriction of $\varphi$ (from $\calA$ to $\calB_+$). We should understand,
however, that to produce nontrivial relations between the integrability
objects, these representations, $\varphi$ and $\rho$, should be somehow
connected. Now, with such a representation of $\calB_+$ in a vector space 
$W$,\footnote{Similarly as before, $W$ is a $\calB_+$-module.}
\[
\rho : \calB_+ \to \End(W),
\]
we introduce the universal $L$-operator
\begin{equation}
\calL_\rho = (\rho \otimes \id) (\calR) \in \End(W) \otimes \calB_-
\label{2.3}
\end{equation}
and the corresponding universal $Q$-operator
\begin{equation}
\calQ_\rho = (\tr_W \otimes \id) (\calL_\rho(\rho(t) \otimes 1)) 
= ((\tr_W \circ \rho) \otimes \id)(\calR (t \otimes 1)),
\label{2.4}
\end{equation}
where the group-like element $t$ is the same as for the universal
transfer-operator, and $1$ again means the unit of the quantum group. 

Further, we can rewrite the master equation (\ref{1.3}) as follows:
\begin{equation}
(\calR^{13} t^1) (\calR^{23} t^2) 
= (\calR^{12})^{-1} (\calR^{23} t^2) (\calR^{13} t^1) 
\calR^{12}. \label{2.5}
\end{equation}
Using the tensor products of the maps $(\tr \circ \varphi_{\zeta})$,
$(\tr \circ \rho_{\zeta})$ for various $\varphi$ and $\zeta$ in the 
providently rewritten master equation (\ref{2.5}), we obtain the first 
functional relations. Applying $(\tr \circ \varphi_{1\zeta_1}) \otimes 
(\tr \circ \varphi_{2\zeta_2})$ to both sides of (\ref{2.5}), we obtain
\begin{equation}
\calT_{\varphi_1}(\zeta_1) \calT_{\varphi_2}(\zeta_2) =
\calT_{\varphi_2}(\zeta_2) \calT_{\varphi_1}(\zeta_1). 
\label{2.6}
\end{equation}
Applying $(\tr \circ \rho_{\zeta_1}) \otimes (\tr \circ \varphi_{\zeta_2})$
to both sides of (\ref{2.5}), we come to the relation
\begin{equation}
\calQ_{\rho}(\zeta_1) \calT_{\varphi}(\zeta_2) =
\calT_{\varphi}(\zeta_2) \calQ_{\rho}(\zeta_1). 
\label{2.7}
\end{equation}
These are the commutativity relations for the universal integrability objects. 

The commutativity between two universal $Q$-operators is not obtained in this
way due to the specific nature of the representation $\rho$, as just described
above. Nevertheless, we have a useful formula for the product of two (and more)
universal $Q$-operators presented here:
\begin{equation}
\calQ_{\rho_1}(\zeta_1) \calQ_{\rho_2}(\zeta_2) = ((\tr{}_{W_1 \otimes W_2} \circ 
(\rho_{1\zeta_1} \otimes_\Delta \rho_{2\zeta_2}) \otimes \id) (\calR (t \otimes 1)),
\label{2.8}
\end{equation}
where we have used the equation
\begin{equation}
\calR^{13} t^1 \calR^{23} t^2 = [(\Delta \otimes \id) (\calR)] [(\Delta \otimes \id) 
(t \otimes 1)] = (\Delta \otimes \id) (\calR (t \otimes 1)).
\label{2.9}
\end{equation} 
Equation (\ref{2.8}) proves to be useful in establishing the commutativity 
between the universal $Q$-operators. However, the very proof of this fact,
and also further functional relations, require more details on the quantum
group and its representations. This is the subject of the next section.

\section{Quantum groups $U_q(\gothgl_n)$ and $U_q(\calL(\gothsl_n))$} \label{s:3}

We need two quantum groups, the quantum group of the general linear Lie group
$\gothgl_n$, and the quantum group of the loop Lie algebra $\calL(\gothsl_n)$,
the latter justifying the term {\em quantum affine algebra\/}.\footnote{The term 
{\em quantum loop algebra\/} is also used.}

Recall that there are $2(n-1)$ generators $E_i$, $F_i$, $i = 1,\ldots,n-1$,
of the Lie algebra $\gothgl_n$, and $n$ basis elements $G_i$, $i = 1,\ldots,n$, 
of its standard Cartan subalgebra $\gothk_n$. They are subject to well-known 
commutation relations and Serre relations. Then, the quantum group 
$U_q(\gothgl_n)$ is generated by the elements\footnote{We use the same notation
for the generators of the quantum group as for the corresponding Lie algebra. 
The notation $q^X$, $X \in \gothk_n$, is used to emphasize that $\gothk_n$
parametrizes the corresponding set of elements of $U_q(\gothgl_n)$.}
\[
E_i, \quad F_i, \quad i = 1,\ldots,n-1, \qquad q^X, \quad X \in \gothk_n,
\]
satisfying the corresponding $q$-deformed defining relations 
\begin{gather}
q^0 = 1, \qquad q^{X_1} q^{X_2} = q^{X_1 + X_2}, \label{3.1} \\
q^X E_i q^{-X} = q^{\alpha_i(X)} E_i, \qquad q^X F_i q^{-X} = q^{-\alpha_i(X)} F_i,
\label{3.2} \\
[E_i, F_j] = \delta_{ij} \kappa_q^{-1}(q^{G_i - G_{i+1}} - q^{G_{i+1} - G_i}),
\label{3.3}
\end{gather}
where $\alpha_i \in \gothk_n^*$ denote the respective simple positive roots, such that
$\alpha_j(G_i) = c_{ij}$, where $c_{ij}$ are the entries of an $n \times (n-1)$ matrix 
with $c_{ii} = 1$, $c_{i+1,i} = - 1$, $i = 1,\ldots,n-1$, and $c_{ij} = 0$ otherwise 
(i. e. for $|j - i| \ge 2$). It means, in particular, that 
$\alpha_j(G_i) - \alpha_j(G_{i+1}) = a_{ij}$ are the entries 
of the Cartan matrix of the Lie algebra $\gothsl_n$. Besides, 
we have the $q$-deformed Serre relations
\begin{gather}
E_i E_j = E_j E_i, \qquad F_i F_j = F_j F_i, \qquad |i - j| \ge 2, \label{3.4} \\
E_i^2 \, E_{i \pm 1} - [2]_q E_i \, E_{i \pm 1} \, E_i + E_{i \pm 1} \, E_i^2 = 0, 
\label{3.5} \\
F_i^2 \, F_{i \pm 1} - [2]_q F_i \, F_{i \pm 1} \, F_i + F_{i \pm 1} \, F_i^2 = 0.
\label{3.6}
\end{gather}
Remember that the deformation parameter $q$ here is the exponential of a complex number
$\hbar$, therefore, the quantum group is just a complex algebra. Here we assume that
\begin{gather*}
q^{X + \nu} = q^\nu q^X, \\
[X + \nu]_q = \kappa_q^{-1} (q^{X + \nu} - q^{-X -\nu}) 
= \kappa_q^{-1} (q^{\nu} q^X - q^{-\nu} q^{-X}) 
\end{gather*}
for any complex number $\nu$ and element $X$ of $\gothk_n$. It is important 
that $U_q(\gothgl_n)$ is a Hopf algebra with respect to appropriately defined 
co-multiplication, antipode and counit; their explicit form can be omitted, 
though.

The generators $E_i$ and $F_i$, $i = 1,\ldots,n-1$, are certainly the root vectors
corresponding to the simple positive and simple negative roots $\alpha_i$ and 
$-\alpha_i$. There are also non-simple roots for $n > 2$, and one constructs 
more root vectors corresponding to these non-simple roots. All these root
vectors are used in constructing the basis vectors of the highest-weight 
$U_q(\gothgl_n)$-modules (see Appendix). Actually, we will need 
infinite-dimensional highest-weight representations 
$\widetilde{\pi}^{\lambda}$ defined by
\begin{equation}
E_i v^\lambda = 0, \quad i = 1,\ldots,n-1, \qquad 
q^X v^\lambda = q^{\lambda(X)} v^\lambda,
\quad X \in \gothk_n, \quad \lambda \in \gothk_n^*, \label{3.7}
\end{equation}
where $v^\lambda$ is the highest-weight vector with the weight $\lambda$ which can 
be seen in terms of its $n$ components $\lambda_i = \lambda(G_i)$. In particular, 
denoting the highest-weight vector by $v_0$, one usually chooses the basis vectors 
for $n = 2$ as
\[
v_k = F^k v_0, \qquad E v_0 = 0, \qquad q^{\nu G_i} v_0 = q^{\nu \lambda_i} v_0, 
\quad i = 1,2,
\]
and for $n = 3$ one chooses
\[
v_k = F_1^{k_1} F_3^{k_3} F_2^{k_2} v_0, \qquad E_i v_0 = 0, \qquad 
q^{\nu G_i} v_0 = q^{\lambda_i} v_0, \quad i = 1,2,3, 
\]
where $F_3 = F_2 F_1 - q F_1 F_2$ is the root vector corresponding to the root 
$- \alpha_1 - \alpha_2$. The infinite-dimensional $U_q(\gothgl_n)$-modules corresponding
to the representations $\widetilde{\pi}^{\lambda}$ are denoted by $\widetilde V^\lambda$.
We will also need the corresponding finite-dimensional representations $\pi^\lambda$ that 
can be obtained from  $\widetilde{\pi}^{\lambda}$ as the quotient representations over 
infinite-dimensional maximal sub-representations in the case when all the differences 
$\lambda_i - \lambda_{i+1}$ are non-\-ne\-ga\-ti\-ve integers. The finite-dimensional
$U_q(\gothgl_n)$-modules corresponding to the representations ${\pi}^{\lambda}$
are denoted by $V^\lambda$. 

The next object, the quantum affine algebra $U_q(\calL(\gothsl_n))$, is more
complicated, because, unlike the preceding case, it is the quantum group of an 
infinite-dimensional Lie algebra. Again, it is a unital associative complex
algebra which can be defined in terms of its generators and Cartan subalgebra.
We start with a reminder that the Lie algebra $\widetilde \calL(\gothsl_n)$ 
has $2n$ generators $e_i$, $f_i$, $i = 0,1,\ldots,n-1$, its Cartan subalgebra 
$\widetilde\gothh_n$ can be described by $n$ basis elements $h_i$, 
$i = 0,1,\ldots,n-1$, and there is a nontrivial center generated 
by the element $c = \sum_{i=0}^{n-1} h_i$. All these elements are 
subject to well-known commutation relations and Serre relations.
The quantum group $U_q(\widetilde\calL(\gothsl_n))$ is generated
by the elements
\[
e_i, \quad f_i, \quad i = 0,1,\ldots,n-1, 
\qquad q^x, \quad x \in \widetilde \gothh_n,
\] 
satisfying certain defining relations. These are the following $q$-deformed
commutation
\begin{gather}
q^0 = 1, \qquad q^{x_1} q^{x_2} = q^{x_1 + x_2}, \label{3.8} \\
q^x e_i q^{-x} = q^{\alpha_i(x)} e_i, \qquad q^x f_i q^{-x} = q^{-\alpha_i(x)} f_i,
\label{3.9} \\
[e_i, f_j] = \delta_{ij} \kappa_q^{-1}(q^{h_i} - q^{-h_i}) \label{3.10}
\end{gather}
and Serre relations
\begin{gather}
\sum_{k = 0}^{1 - \widetilde a_{ij}} (-1)^k \qbinom{1 - \widetilde a_{ij}}{k}_{q} 
(e_i)^{1 - \widetilde a_{ij} - k} e_j (e_i)^k = 0, \label{3.11} \\
\sum_{k = 0}^{1 - \widetilde a_{ij}} (-1)^k \qbinom{1 - \widetilde a_{ij}}{k}_{q} 
(f_i)^{1 - \widetilde a_{ij} - k} f_j (f_i)^k = 0, \label{3.12}
\end{gather}
where $\widetilde a_{ij}$ are the entries of the generalized Cartan matrix 
of type $A^{(1)}_{n-1}$, and the set of simple positive roots is extended 
by the additional root $\alpha_0$.

Since we need finite-dimensional representations, we should first note that there 
is no finite-dimensional representation of $\widetilde \calL(\gothsl_n)$ with 
$c \neq 0$, hence we consider the loop Lie algebra $\calL(\gothsl_n)$ defined 
as the quotient
\[
\calL(\gothsl_n) = \widetilde \calL(\gothsl_n) / \bbC c.
\]
Moreover, also the quantum group $U_q(\widetilde \calL(\gothsl_n))$ does not
have any finite-dimensional representations with $q^{\nu c} \neq 1$ for any
$\nu \in \bbC^\times$. Therefore, we consider the quantum group 
$U_q(\calL(\gothsl_n))$ defined as the quotient
\[
U_q(\calL(\gothsl_n)) = U_q(\widetilde \calL(\gothsl_n)) / 
\langle q^{\nu c} - 1 \rangle_{\nu \in \bbC}.
\]
The quantum group $U_q(\calL(\gothsl_n))$ can be considered in terms of the same
generators and defining relations as $U_q(\widetilde\calL(\gothsl_n))$, where the
additional relation $q^{\nu c} = 1$, $\nu \in \bbC$, is taken into account. The 
restriction $q^{\nu c} = 1$ makes it possible to construct finite-dimensional
representations of the quantum affine algebras under consideration.

The Hopf algebra structure of $U_q(\calL(\gothsl_n))$ can be defined
by the relations with the co-multiplication $\Delta$,
\begin{equation}
\Delta(q^x) = q^x \otimes q^x, \qquad \Delta(e_i) = e_i \otimes 1 + q^{- h_i} \otimes e_i, 
\qquad \Delta(f_i) = f_i \otimes q^{h_i} + 1 \otimes f_i, \label{3.13}
\end{equation}
antipode $S$,
\begin{equation}
S(q^x) = q^{- x}, \qquad S(e_i) = - q^{h_i} e_i, \qquad S(f_i) = - f_i \, q^{- h_i},
\label{3.14}
\end{equation}
and counit $\varepsilon$,
\begin{equation}
\varepsilon(q^x) = 1, \qquad \varepsilon(e_i) = 0, \qquad \varepsilon(f_i) = 0,
\label{3.15}
\end{equation}
acting on the generators explicitly as given.

\section{Jimbo's homomorphism and representations of $U_q(\calL(\gothsl_n))$}
\label{s:4}

As the representation $\varphi$ mentioned in section \ref{s:2} we
use a representation $\widetilde \varphi{}^\lambda_\zeta$ constructed 
as follows.\footnote{In our notation, the tilde corresponds to 
infinite-dimensional representations; for the corresponding 
finite-dimensional representations we use the same symbols simply omitting 
the tilde.} We first define a general grading of the quantum affine algebra $U_q(\calL(\gothsl_n))$ 
with the help of a map $\Gamma_\zeta$ which acts on the generators as
\begin{equation}
\Gamma_\zeta(q^x) = q^x, \quad x \in \widetilde \gothh_n, 
\qquad \Gamma_\zeta(e_i) = \zeta^{s_i} e_i, \qquad 
\Gamma_\zeta(f_i) = \zeta^{-s_i} f_i,
\label{4.1}
\end{equation}
where $\zeta \in \bbC^\times$ is called the {\em spectral parameter\/}, 
and $s_i$ are arbitrary integers. We denote the total sum of these integers by 
$s$. Secondly, we use Jimbo's homomorphism from the quantum affine algebra 
$U_q(\calL(\gothsl_n))$ to the quantum group $U_q(\gothgl_n)$ \cite{Jim86a},
\begin{equation}
\varphi : U_q(\calL(\gothsl_n)) \to U_q(\gothgl_n).
\label{4.2}
\end{equation}
For some values of $n$ of our specific interest, we write it down explicitly. Thus, 
Jimbo's homomorphism $\varphi: U_q(\calL(\gothsl_2)) \to U_q(\gothgl_2)$ is given 
by the relations
\begin{align}
& \varphi(q^{\nu h_0}) = q^{\nu(G_2 - G_1)}, && \varphi(q^{\nu h_1}) = q^{\nu (G_1 - G_2)},
\label{4.3} \\
& \varphi(e_0) = F \, q^{- G_1 - G_2}, && \varphi(e_1) = E, \label{4.4} \\
& \varphi(f_0) = E \, q^{G_1 + G_2} , && \varphi(f_1) = F, \label{4.5} 
\end{align}
and for $n = 3$, we have $\varphi: U_q(\calL(\gothsl_3)) \to U_q(\gothgl_3)$ defined as
\begin{align}
& \varphi(q^{\nu h_0}) = q^{\nu(G_3 - G_1)}, && \varphi(q^{\nu h_1}) = q^{\nu (G_1 - G_2)},
&& \varphi(q^{\nu h_2}) = q^{\nu (G_2 - G_3)}, \label{4.6} \\
& \varphi(e_0) = F_3 \, q^{- G_1 - G_3}, && \varphi(e_1) = E_1, && \varphi(e_2) = E_2, 
\label{4.7} \\
& \varphi(f_0) = E_3 \, q^{G_1 + G_3} , && \varphi(f_1) = F_1, && \varphi(f_2) = F_2,
\label{4.8} 
\end{align} 
where $E_3 = E_1 E_2 - q^{-1} E_2 E_1$ and $F_3 = F_2 F_1 - q F_1 F_2$. Thirdly, and finally, 
we use the highest-weight representation $\widetilde \pi^\lambda$ of $U_q(\gothgl_n)$ briefly 
described in section \ref{s:3}, see equation (\ref{3.7}) and around. We understand that 
Jimbo's homomorphism allows us to use representations of $U_q(\gothgl_n)$ to construct 
representations of $U_q(\calL(\gothsl_n))$. Hence, our basic representation 
$\widetilde \varphi{}^\lambda_\zeta$ is constructed as the superposition of 
the above three maps,
\[
\widetilde \varphi{}^\lambda_\zeta 
= \widetilde \pi^\lambda \circ \varphi \circ \Gamma_\zeta.
\]
Here, using the finite-dimensional representations $\pi^\lambda$ of $U_q(\gothgl_n)$,
we will obtain finite-dimensional representations of $U_q(\calL(\gothsl_n))$ with 
$\varphi{}^\lambda_\zeta = \pi^\lambda \circ \varphi \circ \Gamma_\zeta$. 

Recall again that the universal $R$-matrix is an element of the tensor product
of two Borel subalgebras of the quantum group, in the case under 
consideration, $\calR \in U_q(\gothb_+) \otimes U_q(\gothb_-)$, where the Borel
subalgebra $U_q(\gothb_+)$ of $U_q(\calL(\gothsl_n))$ is generated by the elements
$e_i$, $i = 0,1,\ldots,n-1$, and $q^x$, and the Borel subalgebra $U_q(\gothb_-)$ of 
$U_q(\calL(\gothsl_n))$ is generated by the elements $f_i$, $i = 0,1,\ldots,n-1$, 
and $q^x$. One can try to define the representation
$\rho$ of $U_q(\gothb_+)$ introduced in section \ref{s:2} in the following 
way. Let $\widetilde \varphi{}^\lambda_\zeta$ be a representation of 
$U_q(\calL(\gothsl_n))$ constructed in accordance with the above
prescription, and $\xi \in \widetilde \gothh_n^*$. Then the relations
\begin{equation}
\widetilde \varphi{}^\lambda_\zeta[\xi](e_i) = \widetilde \varphi{}^\lambda_\zeta(e_i),
\qquad \widetilde \varphi{}^\lambda_\zeta[\xi](q^x) 
= q^{\xi(x)} \widetilde \varphi{}^\lambda_\zeta(q^x) 
\label{4.9}
\end{equation}
define a representation of $U_q(\gothb_+)$ called a {\em shifted representation\/}.
We see that the only difference between the shifted and initial representations   
appears simply in the factor $q^{\xi(x)}$ in the second relation of (\ref{4.9}). 
Moreover, choosing the twist element $t$ explicitly in the form
\[
t = q^{\sum_{i=0}^{n-1} \phi_i h_i / n}
\]
for some complex numbers $\phi_i$, which are subject to the condition $\sum_i \phi_i = 0$ 
to respect the restriction $q^{\nu c} = 1$, we obtain that the universal integrability 
objects based on the representation $\widetilde \varphi{}^\lambda_\zeta$ are related to 
the universal integrability objects based on its shifted counterpart 
$\widetilde \varphi{}^\lambda_\zeta[\xi]$ simply as
\begin{equation}
\calT_{\widetilde{\varphi}{}^\lambda[\xi]}(\zeta) = \calT_{\widetilde{\varphi}{}^\lambda}(\zeta) 
\, q^{\sum_{i=0}^{n-1}\xi(h_i) h'_i / n}, \label{4.10}
\end{equation}
where we use the notation $h'_i = h_i + \phi_i$. 

It is not difficult to see that the shifted representation $\widetilde{\varphi}{}^\lambda_\zeta[\xi]$ 
for a nonzero $\xi$ cannot be extended to a representation of the full quantum affine algebra. Therefore, the shifted representation is what we actually need to construct analogs of the representation $\rho$ introduced in section \ref{s:2}. It is clear that the universal $Q$-operators constructed with the help of the shifted representations $\widetilde{\varphi}{}^\lambda_\zeta[\xi]$ will be connected with the corresponding universal transfer operators based on $\widetilde{\varphi}{}^\lambda_\zeta$ according to (\ref{4.10}). Hence, we do not obtain a really new object. However, considering 
all the $U_q(\gothb_+)$-modules with general nonzero shifts $\xi$, we can demonstrate that there are interesting limits of the corresponding representations when the differences $\lambda_i - \lambda_{i+1} = \mu_i$, $i = 1,\ldots,n-1$, go to infinity and $\xi$ is chosen appropriately.  To be precise, we note that for $n > 2$ the obtained representation is reducible and we take the corresponding irreducible subrepresentation. The final formulas 
for the $Q$-operators look for $n = 2$ as
\begin{equation*}
\calQ(\zeta) = \lim_{\mu \to \infty} \left( \widetilde \calT^{(\mu, 0)} 
(q^{- 1 / s} \zeta) \, q^{(\mu h'_0 - \mu h'_1) / 2} \right),
\end{equation*}
and for $n = 3$ as
\begin{equation*}
\calQ(\zeta) = (1 - q^{(h_0' - 2 h_1' + h_2')/3}) \lim_{\mu_1, \mu_2 \to \infty} 
\left( \widetilde \calT^{(\mu_1 + \mu_2, \mu_2, 0)} 
(q^{- 2 / s} \zeta) \, q^{((\mu_1 + \mu_2) h'_0 - \mu_1 h'_1 - \mu_2 h'_2) / 3} \right).
\end{equation*}
Here, we use the notation $\widetilde \calT^\lambda(\zeta)$ for $\calT_{\widetilde \varphi^\lambda}(\zeta)$, and we do not write explicitly the representation $\rho_\zeta$ 
in $\calQ(\zeta)$, simply keeping in mind that it is obtained from $\widetilde \varphi^\lambda_\zeta[\xi]$ by the procedure shortly described above. 

The above two (lower and higher rank) basic examples can be directly generalized 
to the case of general $n$. Remarkably, it follows from the limit relation between 
the universal integrability objects that also the universal $Q$-operators commute as 
well as the universal transfer operators in (\ref{2.6}) and (\ref{2.7}),
\begin{equation}
\calQ_{\rho_1}(\zeta_1) \calQ_{\rho_2}(\zeta_2) = \calQ_{\rho_2}(\zeta_2) \calQ_{\rho_1}(\zeta_1).
\label{4.13}
\end{equation}

The representations $\rho_\zeta$ have a useful interpretation in terms of the so-called {\em $q$-oscillators\/},
see e. g. \cite{KliSch97, BazLukZam97, BazLukZam99, BazHibKho02}. We define them by a natural deformation of the 
usual oscillators with the deformation parameter $q = \exp\hbar$, where $\hbar$ is a complex number, such that $q$ 
is not a root of unity. The $q$-oscillator algebra $\Osc_q$ is a unital associative $\bbC$-algebra with generators 
$b^\dagger$, $b$, $q^{\nu N}$, $\nu \in \bbC$, and relations
\begin{gather*}
q^0 = 1, \qquad q^{\nu_1 N} q^{\nu_2 N} = q^{(\nu_1 + \nu_2)N}, \\
q^{\nu N} b^\dagger q^{-\nu N} = q^\nu b^\dagger, \qquad q^{\nu N} b q^{-\nu N} = q^{-\nu} b, \\
b^\dagger b = [N]_q, \qquad b b^\dagger = [N + 1]_q.
\end{gather*}
The monomials $(b^\dagger)^{k + 1} q^{\nu N}$, $b^{k + 1} q^{\nu N}$ and 
$q^{\nu N}$ for $k \in \bbZ_+$ and $\nu \in \bbC$ form a basis of $\Osc_q$.

The representations of the $q$-oscillator algebra are constructed as follows. 
One can see that the relations
\begin{gather}
q^{\nu N} v_k = q^{\nu k} v_k, \label{osca} \\*
b^\dagger v_k = v_{k + 1}, \qquad b \, v_k = [k]_q v_{k - 1}, \label{oscb}
\end{gather}
supplied with the assumption $v_{-1} = 0$, endow free vector space generated by the 
set $\{ v_0, v_1, \ldots \}$ with the structure of an $\Osc_q$-module. We denote the corresponding representations by $\chi$.\footnote{There is one more useful representation, 
but we do not use it here.}

In the case of $n = 2$ for the representation $\rho_\zeta$ we have
\begin{align*}
& q^{\nu h_0} v_k = q^{2 \nu k} v_k, & & q^{\nu h_1} v_k = q^{- 2 \nu k} v_k, \\
& e_0 v_k = v_{k + 1}, & & e_1 v_k = \kappa_q^{-1} q^{- k} [k]_q v_{k - 1}.
\end{align*}
Here $v_k$, $k \in \bbZ_+$, are the basis vectors in the representation space. Comparing these relations with (\ref{osca}) and (\ref{oscb}), we see that it is natural to define the mapping $\theta$ from $\uqbp$ to $\Osc_q$ as
\begin{align*}
& \theta(q^{\nu h_0}) = q^{2 \nu N}, && \theta(q^{\nu h_1}) = q^{- 2\nu N}, \\*
& \theta(e_0) = b^\dagger, && \theta(e_1) = \kappa_q^{-1} b \, q^{- N}.
\end{align*}
It can be shown that $\theta$ is a homomorphism. Now, for the representation $\rho_\zeta$ we have
\begin{equation}
\rho_\zeta = \chi \circ \theta \circ \Gamma_\zeta. \label{rhoz}
\end{equation}
In the case of $n = 3$ we need two copies of the algebra $\Osc_q$. The corresponding homomorphism from $\uqbp$ to $\Osc_q \otimes \Osc_q$ has here the form
\begin{align*}
& \theta(q^{\nu h_0}) = q^{\nu(2 N_1 + N_2)}, && \theta(q^{\nu h_1}) = q^{\nu(- N_1 + N_2)}, &&
\theta(q^{\nu h_2}) = q^{\nu(- N_1 - 2 N_2)}, \\*
& \theta(e_0) = b^\dagger_1 q^{- N_2}, && \theta(e_1) = - b_1^{\mathstrut} b_2^\dagger q^{- N_1 + N_2 + 1}, && \theta(e_2) = \kappa_q^{-1} b_2^{\mathstrut} q^{- N_2}.
\end{align*}

One can produce more universal integrability objects with the help of the automorphisms 
of the quantum group $U_q(\calL(\gothsl_n))$. These are the automorphism $\sigma$ 
corresponding to the cyclic permutations of the Dynkin diagram of the Kac--Moody Lie 
algebra of type $A^{(1)}_{n-1}$ transforming the simple positive roots as $\alpha_i \to \alpha_{i+1}$, and the automorphism $\tau$ acting as $\alpha_i \to \alpha_{n-i}$ 
while leaving $\alpha_0$ alone. Here, $\sigma^n = \id$ and $\tau^2 = \id$.
If $\theta : U_q(\gothb_+) \to \underbrace{\Osc_q \otimes \cdots \otimes \Osc_q}_{n - 1}$ is the initial homomorphism for the basic representation $\rho_\zeta$, we define
\[
\theta_i = \theta \circ \sigma^{-i}, \qquad \overline\theta_i = \theta \circ \tau \circ \sigma^{-i+1}, \qquad i = 1,\ldots,n,
\]
and use the formulas of type (\ref{rhoz}) to define the representations $\rho_{i\zeta}$ and $\ovrho_{i\zeta}$ and then the set of $2 n$ universal $Q$-operators. Similarly, one can 
produce more representations $\widetilde \varphi^\lambda_{i\zeta}$ with the help of the 
automorphisms $\sigma$ and $\tau$ starting from a basic one, and construct the respective universal transfer operators. However, not all of them will be independent. In fact, there 
is only one universal transfer operator for $n = 2$, and there are two independent universal transfer operators in the higher-rank case.

\section{The universal functional relations} \label{s:5}

\subsection{The key relations}

Thus, we have $2n$ different universal $Q$-operators.\footnote{One has only two universal $Q$-operators for 
the lower-rank case (for $n=2$).} Specifying the irreducible representations $\rho_\zeta$ and constructing the 
corresponding universal $Q$-operators, we can use the formula for their products given in section \ref{s:2}. We see here that to analyze these products, we have to consider tensor products $(\rho_{\ell\zeta_\ell} \otimes_\Delta \cdots \otimes_\Delta \rho_{1\zeta_1})$ of the representations $\rho_{i \zeta}$, with $\ell = 2,3,\ldots,n$, and go similarly with $\ovrho_{i\zeta}$. Then, choosing appropriate bases for the 
corresponding $U_q(\gothb_+)$-modules $(W_\ell)_{\zeta_\ell} \otimes_\Delta \cdots \otimes_\Delta (W_1)_{\zeta_1}$, 
we should consider their defining module relations. In this way we obtain the whole set of the universal functional 
relations.

In particular, putting $\ell = 2$ in a higher-rank case ($n = 3$), we see that the $U_q(\gothb_+)$-module $(W_2)_{\zeta_2} \otimes_\Delta (W_1)_{\zeta_1}$ for some special choice of $\zeta_1$ and $\zeta_2$ is reducible and the corresponding quotient module is isomorphic to $(\ovW_3)_{\zeta}[\xi]$ with a new spectral parameter $\zeta$ expressed in terms of $\zeta_1$ and $\zeta_2$, and a certain shift $\xi$ of the corresponding representation $\ovrho_{3 \zeta}$. Similarly, one can obtain the $U_q(\gothb_+)$-module $(W_3)_{\zeta}[\xi]$ for some $\zeta$ and $\xi$ as a quotient of $(\ovW_2)_{\zeta_2} \otimes_\Delta (\ovW_1)_{\zeta_1}$. 
This consideration allows us to write down the universal double-$Q$ functional relations 
in the determinant form,
\begin{gather}
\calC_i \ovcalQ_i(\zeta) = \calQ_j(q^{1/s} \zeta) \calQ_k(q^{- 1/s} \zeta) 
- \calQ_j(q^{-1/s} \zeta) \calQ_k(q^{1/s} \zeta), 
\label{5.1}\\
\calC_i \calQ_i(\zeta) = \ovcalQ_j(q^{- 1/s} \zeta) \ovcalQ_k(q^{1/s} \zeta) 
- \ovcalQ_j(q^{1/s} \zeta) \ovcalQ_k(q^{- 1/s} \zeta),
\label{5.2}
\end{gather}
where $(i,j,k)$ runs over all cyclic permutations of the set $(1,2,3)$, and we use the notation
\begin{equation}
\calC_i = q^{- \calD_i / s} (q^{2 \calD_j / s} - q^{2 \calD_k / s})^{-1}, \qquad 
\calD_i = (h'_{i-1} - h'_i) s / 6. \label{5.3}
\end{equation}
Note that such relations are absent in the lower-rank case.

To derive the major functional relations, we have to put $\ell = n$ and analyze the $n$-tuple
product representation $(\rho_{n\zeta_n} \otimes_\Delta \cdots \otimes_\Delta \rho_{1\zeta_1})$
corresponding to the tensor product of $n$ $U_q(\gothb_+)$-modules $(W_n)_{\zeta_n} \otimes_\Delta 
\cdots \otimes_\Delta (W_1)_{\zeta_1}$. It allows one to obtain the key relation between the universal
transfer operator and universal $Q$-operators,
\begin{equation}
\calC \, \widetilde \calT^\lambda(\zeta) = \calQ_1(q^{- 2 (\lambda + \rho)_1/s} \zeta) 
\cdots  \calQ_n(q^{- 2 (\lambda + \rho)_n/s} \zeta), \qquad \calC = \calC_1 \cdots \calC_n,
\label{5.4}
\end{equation} 
where $\rho \in \gothk_n^*$ is the half-sum of the positive roots,
\[
\rho = ((n-1)/2, (n-3)/2, \ldots, -(n-1)/2).
\]
This is the central equation from which follow all other functional relations. One can see that 
relation (\ref{5.4}) contains only the universal transfer operator for the infinite-dimensional 
representation $\widetilde \varphi^\lambda_\zeta$. To obtain the corresponding relations for the
universal transfer operator based on the finite-dimensional representation $\varphi^\lambda_\zeta$,
one uses the quantum version of the so-called {\em B.G.G. resolution\/} \cite{Ros91, BooGoeKluNirRaz14b, 
NirRaz14}, which implies an exact sequence of $U_q(\gothgl_n)$-modules and $U_q(\gothgl_n)$-homomorphisms,
\begin{gather*}
U_k = \bigoplus_{\substack{w \in W \\ \ell(w) = k}} \widetilde V^{w \cdot \lambda},
\qquad w \cdot \lambda = w(\lambda + \rho) - \rho, \\
0 \longrightarrow U_n \overset{\varphi_n}{\longrightarrow} U_{n-1} \overset{\varphi_{n-1}}{\longrightarrow}
\ldots \overset{\varphi_1}{\longrightarrow} U_0 \overset{\varphi_0}{\longrightarrow} U_{-1} \longrightarrow 0, 
\qquad U_{-1} = V^\lambda,
\end{gather*}
where $w$ means any element of the Weyl group $W$ of the root system of $\gothgl_n$, with $\ell(w)$
being its length, and $w \cdot \lambda$ stands for the affine action of $w$ defined explicitly as above. 
The B.G.G. resolution allows one to express the trace over the finite-dimensional $U_q(\gothgl_n)$-module 
$V^\lambda$ as certain linear combination of traces over the infinite-dimensional $U_q(\gothgl_n)$-modules
$\widetilde V^{w \cdot \lambda}$,
\begin{equation}
\tr^\lambda = \sum_{w \in W} (-1)^{\ell(w)} \widetilde \tr^{w \cdot \lambda}.
\label{5.5}
\end{equation}
Then, using our definition of the universal transfer operator (\ref{2.2}), we immediately 
come to a remarkable relation between the universal transfer operator based on the finite-dimensional representation $\varphi^\lambda_\zeta$ and universal transfer 
operators based on the infinite-dimensional representations 
$\widetilde \varphi^{w \cdot \lambda}_\zeta$,
\begin{equation}
\calT^\lambda(\zeta) = \sum_{p \in \mathrm S_n} \sgn(p) \, \widetilde \calT^{\, p(\lambda + \rho) - \rho}(\zeta).
\label{5.6}
\end{equation}
It is taken into account here that $W$ can be identified with the symmetric group 
$S_n$, and so, $p$ is an element of $S_n$ acting on $\gothk_n^*$ by appropriate 
permutations. For example, one obtains 
\[
\calT^{(\lambda_1 , \, \lambda_2)}(\zeta) = \widetilde \calT^{(\lambda_1 , \, \lambda_2)}(\zeta) - \widetilde \calT^{(\lambda_2 - 1 , \, \lambda_1 + 1)}(\zeta)
\]
in the lower-rank case ($ n = 2$), and 
\begin{multline*}
\calT^{(\lambda_1, \, \lambda_2, \, \lambda_3)}(\zeta) = \widetilde \calT^{(\lambda_1, \, \lambda_2, \, \lambda_3)}(\zeta) - \widetilde \calT^{(\lambda_2 - 1, \, \lambda_1 + 1, \, \lambda_3)}(\zeta) - \widetilde \calT^{(\lambda_1, \, \lambda_3 - 1, \, \lambda_2 + 1)}(\zeta) \\ + \widetilde \calT^{(\lambda_3 - 2, \, \lambda_1 + 1, \, \lambda_2 + 1)}(\zeta)
+ \widetilde \calT^{(\lambda_2 - 1, \, \lambda_3 - 1, \, \lambda_1 + 2)}(\zeta)
- \widetilde \calT^{(\lambda_3 - 2, \, \lambda_2, \, \lambda_1 + 2)}(\zeta)
\end{multline*}
in the higher-rank case ($n = 3$). Relation (\ref{5.6}) leads to the following 
representation of the universal transfer operator in terms of the universal 
$Q$-operators:
\begin{equation}
\calC \, \calT^{\lambda - \rho}(\zeta) = \det \left( \calQ_i(q^{- 2 \lambda_j / s} \zeta) \right)_{i,j = 1,\ldots,n}.
\label{5.7}
\end{equation}
Similar consideration holds as well when also the automorphism $\tau$ is involved giving 
rise to the barred universal integrability objects. In this case we come to the determinant representation
\begin{equation}
\calC \, \ovcalT^{\lambda - \rho}(\zeta) 
= \det \left( \ovcalQ_i(q^{2 \lambda_j /s } \zeta) \right)_{i,j = 1,\ldots,n}.
\label{5.8}
\end{equation}
It is worthwhile noting that for the lower-rank case, $n = 2$, we would be able to obtain 
the universal functional relations in the determinant form even if we started with Jimbo's 
homomorphism from $U_q(\calL(\gothsl_2))$ to $U_q(\gothsl_2)$. However, for higher-rank 
cases, $n \ge 3$, it is most convenient to use Jimbo's homomorphism from 
$U_q(\calL(\gothsl_n))$ to $U_q(\gothgl_n)$.

\subsection{Universal $TQ$- and $TT$-relations}

Further universal functional relations can be obtained from the vanishing of the determinants of certain 
$(n+1) \times (n+1)$ matrices with one dependent row using equations (\ref{5.7}), (\ref{5.8}). The universal 
$TQ$-relations follow from the identity
\begin{equation*}
\det \left( 
\begin{array}{cccc}
\calQ_1(q^{-2\lambda_1/s} \zeta) & \cdots & \calQ_1(q^{-2\lambda_n/s} \zeta) &
\calQ_1(q^{-2\lambda_{n+1}/s} \zeta) \\
\calQ_2(q^{-2\lambda_1/s} \zeta) & \cdots & \calQ_2(q^{-2\lambda_n/s} \zeta) &
\calQ_2(q^{-2\lambda_{n+1}/s} \zeta) \\
\vdots & \ddots & \vdots & \vdots \\
\calQ_n(q^{-2\lambda_1/s} \zeta) & \cdots & \calQ_n(q^{-2\lambda_n/s} \zeta) &
\calQ_n(q^{-2\lambda_{n+1}/s} \zeta) \\
\calQ_j(q^{-2\lambda_1/s} \zeta) & \cdots & \calQ_j(q^{-2\lambda_n/s} \zeta) &
\calQ_j(q^{-2\lambda_{n+1}/s} \zeta) 
\end{array}
\right) = 0
\end{equation*}
obviously satisfied for any $j = 1,\ldots,n$. Such matrices can be constructed also for the 
barred universal integrability objects. In the lower-rank case, putting $n = 2$, we obtain
\begin{multline}
\calT^{(\lambda_1 - 1/2, \, \lambda_2 + 1/2)}(\zeta) \calQ_j(q^{- 2 \lambda_3 / s} \zeta) 
\\ - \calT^{(\lambda_1 - 1/2, \, \lambda_3 + 1/2)}(\zeta) \calQ_j(q^{- 2 \lambda_2 / s} \zeta)
\\ + \calT^{(\lambda_2 - 1/2, \, \lambda_3 + 1/2)}(\zeta) \calQ_j(q^{- 2 \lambda_1 / s} \zeta) = 0.
\label{5.9}
\end{multline}
We call this equation the universal $TQ$-relation. For example, choosing in (\ref{5.9}) 
the components of $\lambda$ as $\lambda_1 = 1$, $\lambda_2 = 0$, $\lambda_3 = -1$, after some 
transformations based on obvious symmetry properties of $\calT^{(\lambda_1,\lambda_2)}(\zeta)$, 
we derive
\begin{equation*}
\calT^{(1/2, \, -1/2)}(\zeta) \calQ_j(\zeta) = \calQ_j(q^{2/s} \zeta) +  \calQ_j(q^{-2/s} \zeta).
\end{equation*}
This is the universal analog of the famous Baxter's $TQ$-relation, that is given in the form 
independent of the representation of the quantum group in the quantum space.

In a higher-rank case, $n = 3$, we obtain the $U_q(\calL(\gothsl_3))$ universal $TQ$-relations
\begin{multline}
\calT^{(\lambda_1 - 1, \, \lambda_2, \, \lambda_3 + 1)}(\zeta) \calQ_j(q^{- 2 \lambda_4 / s} \zeta) 
- \calT^{(\lambda_1 - 1, \, \lambda_2, \, \lambda_4 + 1)}(\zeta) \calQ_j(q^{- 2 \lambda_3 / s} \zeta) \nonumber \\
+ \calT^{(\lambda_1 - 1, \, \lambda_3, \, \lambda_4 + 1)}(\zeta) \calQ_j(q^{- 2 \lambda_2 / s} \zeta) 
- \calT^{(\lambda_2 - 1, \, \lambda_3, \, \lambda_4 + 1)}(\zeta) \calQ_j(q^{- 2 \lambda_1 / s} \zeta) = 0.
\label{5.10}
\end{multline}
Now, with the choice $\lambda_1 = 2$, $\lambda_2 = 1$, $\lambda_3 = 0$, $\lambda_4 = -1$, also using symmetry 
properties of $\calT^{(\lambda_1,\lambda_2,\lambda_3)}(\zeta)$, we derive 
\begin{equation}
\calT^{(1, \, 1, \, 0)}(\zeta) \calQ_j(\zeta)
- \calT^{(1, \, 0, \, 0)}(\zeta) \calQ_j(q^{- 2 / s} \zeta) = 
\calQ_j(q^{2 / s} \zeta)  -  \calQ_j(q^{- 4 / s} \zeta).
\label{5.11}
\end{equation}
In a similar way we obtain
\begin{equation}
\ovcalT^{(1, \, 1, \, 0)}(\zeta) \ovcalQ_j(\zeta) 
- \ovcalT^{(1, \, 0, \, 0)}(\zeta) \ovcalQ_j(q^{2 / s} \zeta) =
\ovcalQ_j(q^{- 2 / s} \zeta)  -  \ovcalQ_j(q^{4 / s} \zeta). 
\label{5.12}
\end{equation}
Unlike the lower-rank case, each of the above equations involves different universal transfer operators. However,
one can derive functional relations containing only one universal transfer operator, $\calT^{(1, \, 0, \, 0)}(\zeta)$
or $\calT^{(1, \, 1, \, 0)}(\zeta)$, or their barred analog, but having mixed $\calQ_i$ and $\ovcalQ_j$ for distinct
$i$ and $j$ in one same equation. To this end, one can use the Jacoby identity for determinants \cite{Hir04}. Then, 
in the simplest higher-rank case $n = 3$, we obtain the corresponding $TQ\overline{Q}$-relations
\begin{multline*}
\calT^{(1,0,0)}(\zeta) \calQ_i(q^{- 2/s}\zeta) {\ovcalQ}_j(q^{- 1 / s} \zeta) =   
\calQ_i(q^{-4/s}\zeta) {\ovcalQ}_j(q^{- 1 /s} \zeta) 
\\ + \calQ_i(\zeta) {\ovcalQ}_j(q^{- 3/s}\zeta) + \calQ_i(q^{- 2/s}\zeta) {\ovcalQ}_j(q^{1/s}\zeta)
\end{multline*}
and
\begin{multline*}
\calT^{(1,1,0)}(\zeta) \calQ_i(\zeta) {\ovcalQ}_j(q^{-1/s}\zeta)
= \calQ_i(q^{2/s}\zeta) {\ovcalQ}_j(q^{-1/s}\zeta) \\ 
+ \calQ_i(q^{-2/s}\zeta) {\ovcalQ}_j(q^{1/s}\zeta) + \calQ_i(\zeta) {\ovcalQ}_j(q^{-3/s}\zeta). 
\end{multline*}

The universal $TT$-relations can be derived from the equation
\begin{equation*}
\det \left( 
\begin{array}{cccc}
\calQ_1(q^{-2\lambda_1/s} \zeta) & \cdots & \calQ_1(q^{-2\lambda_n/s} \zeta) &
\calQ_1(q^{-2\lambda_{n+1}/s} \zeta) \\
\calQ_2(q^{-2\lambda_1/s} \zeta) & \cdots & \calQ_2(q^{-2\lambda_n/s} \zeta) &
\calQ_2(q^{-2\lambda_{n+1}/s} \zeta) \\
\vdots & \ddots & \vdots & \vdots \\
\calQ_n(q^{-2\lambda_1/s} \zeta) & \cdots & \calQ_n(q^{-2\lambda_n/s} \zeta) &
\calQ_n(q^{-2\lambda_{n+1}/s} \zeta) \\
\calT^{(\lambda'_1,\lambda'_{n+2},\ldots,\lambda'_{2n})}(\zeta) & \cdots &
\calT^{(\lambda'_n,\lambda'_{n+2},\ldots,\lambda'_{2n})}(\zeta)
& \calT^{(\lambda'_{n+1},\lambda'_{n+2},\ldots,\lambda'_{2n})}(\zeta) 
\end{array}
\right) = 0, 
\end{equation*}
where $\lambda'_j = \lambda_j - (n-1)/2$, $j = 1,\ldots,n+1$, and $\lambda'_{n+1+k} = \lambda_{n+1+k} - (n-2k-1)/2$,
$k = 1,\ldots,n-1$. Using equation (\ref{5.7}) to express the operators $\calT^{(\lambda'_j,\lambda'_{n+2}, \ldots, 
\lambda'_{2n})}(\zeta)$ in terms of the universal $Q$-operators for all $j = 1,\ldots,n+1$, one can see that the last 
row of the above $(n+1) \times (n+1)$ matrix is a linear combination of the first $n$ rows. Therefore, its determinant 
identically vanishes. Expanding this determinant over the last row and using again (\ref{5.7}), we obtain the universal 
$TT$-relations. The case with the barred universal integrability objects can be considered in the same way. Actually,
the universal $TT$-relations for the barred integrability objects can be obtained directly from the relations for the 
unbarred quantities changing $q$ by $q^{-1}$ there \cite{BooGoeKluNirRaz14b}. 

In the simplest lower-rank case we obtain
\begin{multline}
\calT^{(\lambda_1 - 1/2, \, \lambda_2 + 1/2)}(\zeta) \calT^{(\lambda_3 - 1/2, \, \lambda_4 + 1/2)}(\zeta) 
\\ - \calT^{(\lambda_1 - 1/2, \, \lambda_3 + 1/2)}(\zeta) \calT^{(\lambda_2 - 1/2, \, \lambda_4 + 1/2)}(\zeta) 
\\ + \calT^{(\lambda_2 - 1/2, \, \lambda_3 + 1/2)}(\zeta) \calT^{(\lambda_1 - 1/2, \, \lambda_4 + 1/2)}(\zeta) 
= 0. \label{5.13}
\end{multline}
Choosing the weights subsequently as $\lambda_1 = \ell + 1$, $\lambda_2 = \ell$, $\lambda_3 = 0$, 
$\lambda_4 = - 1$ and $\lambda_1 = \ell + 1$, $\lambda_2 = \ell$, $\lambda_3 = \ell - 1$, $\lambda_4 = - 1$, we 
derive the universal analog of the $TT$-relations of particular interest,  
\begin{equation}
\calT^{(\ell,0)}(q^{-1/s}\zeta) \calT^{(\ell,0)}(q^{1/s} \zeta) = 1 + 
\calT^{(\ell-1,0)}(q^{-1/s}\zeta) \calT^{(\ell+1,0)}(q^{1/s} \zeta) \label{5.14}
\end{equation}
and 
\begin{equation}
\calT^{(1,0)}(q^{-2\ell/s}\zeta) \calT^{(\ell,0)}(\zeta) = \calT^{(\ell+1,0)}(\zeta) + 
\calT^{(\ell-1,0)}(\zeta),  \label{5.15}
\end{equation}
respectively.

For a more complicated higher-rank case, $n = 3$, we obtain
\begin{multline}
\calT^{(\lambda_1 - 1, \, \lambda_2, \, \lambda_3 + 1)}(\zeta) 
\calT^{(\lambda_4 - 1, \, \lambda_5, \, \lambda_6 + 1)}(\zeta) 
- \calT^{(\lambda_1 - 1, \, \lambda_2, \, \lambda_4 + 1)}(\zeta) 
\calT^{(\lambda_3 - 1, \, \lambda_5, \, \lambda_6 + 1)}(\zeta) \\
 + \calT^{(\lambda_1 - 1, \, \lambda_3, \, \lambda_4 + 1)}(\zeta) 
\calT^{(\lambda_2 - 1, \, \lambda_5, \, \lambda_6 + 1)}(\zeta) 
- \calT^{(\lambda_2 - 1, \, \lambda_3, \, \lambda_4 + 1)}(\zeta) 
\calT^{(\lambda_1 - 1, \, \lambda_5, \, \lambda_6 + 1)}(\zeta) = 0. \label{5.16}
\end{multline}
Putting here $\lambda_1 = \ell + 2$, $\lambda_2 = \ell + 1$, $\lambda_3 = 1$, 
$\lambda_4 = 0$, $\lambda_5 = 0$, $\lambda_6 = -1$, we derive one $TT$-relation 
of particular interest in the universal form, 
\begin{equation}
\calT^{(\ell - 1, \, 0, \, 0)}(\zeta) \calT^{(\ell + 1, \, 0, \, 0)}(q^{2/s} \zeta) 
= \calT^{(\ell, \, 0, \, 0)}(\zeta) \calT^{(\ell, \, 0, \, 0)}(q^{2/s} \zeta) 
- \calT^{(\ell, \, \ell, \, 0)}(\zeta), \label{5.17}
\end{equation}
while the choice $\lambda_1 = \ell + 2$, $\lambda_2 = \ell$, $\lambda_3 = 0$, 
$\lambda_4 = \ell + 1$, $\lambda_5 = \ell + 1$, $\lambda_6 = -1$ produces another 
interesting universal $TT$-relation,
\begin{equation}
\calT^{(\ell - 1, \, \ell - 1, \, 0)}(q^{-2/s} \zeta) \calT^{(\ell + 1, \, \ell + 1, \, 0)}(\zeta) 
= \calT^{(\ell, \, \ell, \, 0)}(q^{-2/s} \zeta) \calT^{(\ell, \, \ell, \, 0)}(\zeta) - 
\calT^{(\ell, \, 0, \, 0)}(\zeta). \label{5.18}
\end{equation}
Relations of the type (\ref{5.14}), (\ref{5.15}) and (\ref{5.17}), (\ref{5.18}) are usually called 
fusion relations, see \cite{KluPea92, KunNakSuz94, BazHibKho02}. They are given here in the form 
independent of the representation of the quantum group in the quantum space, that is, they are the 
{\em universal fusion relations\/}.

\subsection{Quantum Jacobi--Trudi identity}

Equation (\ref{5.15}) allows one to express $\calT^{(\ell,0)}$ through $\calT^{(1,0)}$. In the higher-rank
case the situation is more intricate. Considering $n = 3$, we first note that equation (\ref{5.17}) can be
generalized as follows. Let $\ell_1$, $\ell_2$ be positive integers, such that $\ell_1 \ge \ell_2$. Choosing 
the weight components $\lambda_i$ in (\ref{5.16}) as  $\lambda_1 = \ell_1 + 2$, $\lambda_2 = \ell_2 + 1$, 
$\lambda_3 = 1$, $\lambda_4 = 0$, $\lambda_5 = 0$, $\lambda_6 = -1$, we obtain
\begin{equation}
\calT^{(\ell_1, \, \ell_2, \, 0)}(\zeta) = \calT^{(\ell_1, \, 0, \, 0)}(\zeta) 
\calT^{(\ell_2, \, 0, \, 0)}(q^{2/s} \zeta) - \calT^{(\ell_1 + 1, \, 0, \, 0)}(q^{2/s} \zeta) 
\calT^{(\ell_2 - 1, \, 0, \, 0)}(\zeta). \label{5.19}
\end{equation}
Equation (\ref{5.19}) implies that $\calT^{(\ell_1,\ell_2,0)}(\zeta)$ can be expressed via 
$\calT^{(\ell,0,0)}(\zeta)$, and subsequently via $\calT^{(1,0,0)}(\zeta), \, \calT^{(1,1,0)}(\zeta)$.
Explicitly, the result of this procedure is given by the determinant 
\begin{equation}
\calT^{(\ell_1, \, \ell_2, \, 0)}(\zeta) 
= \det \left( \calE_{\ell^t_i - i + j}(q^{- 2(j - 1)/s} \zeta) 
\right)_{1 \le i, \, j \le \ell_1}, \label{5.20}
\end{equation}
where we use the notation
\begin{gather*}
\calE_0(\zeta) = \calE_3(\zeta) = 1, \qquad 
\calE_1(\zeta) = \calT^{(1, \, 0, \, 0)}(\zeta), \qquad 
\calE_2(\zeta) = \calT^{(1, \, 1, \, 0)}(\zeta), \\
\calE_k(\zeta) = 0 \qquad \forall \; k < 0, \; k > 3, \\
\ell^t_i = 2, \quad 1 \le i \le \ell_2, \qquad \ell^t_i = 1, \quad \ell_2 < i \le \ell_1.
\end{gather*}
Here, one connects the integers $\ell_1$ and $\ell_2$ with the Young diagram with the rows 
of the length $\ell_1$ and $\ell_2$, then the numbers $\ell_i^t$ describe the rows of the 
transposed diagram. Equation (\ref{5.20}) can thus be regarded as the universal quantum 
analog of the Jacobi--Trudi identity from the theory of symmetric polynomials \cite{Che87, 
BazRes90, BazHibKho02}. To prove this identity in the universal form, we use $(n+1)$-term 
universal functional relations and certain symmetries of the universal transfer operators 
\cite{BooGoeKluNirRaz14b}.

Remarkably, also the barred universal transfer operators $\ovcalT^{(\ell_1,\ell_2,0)}(\zeta)$ 
can be expressed through the same basic universal transfer operators $\calT^{(1,0,0)}(\zeta)$ 
and $\calT^{(1,1,0)}(\zeta)$. To come to this conclusion, one can write the Jacobi--Trudi 
identity for $\ovcalT^{(\ell_1,\ell_2,0)}(\zeta)$ down and use the relations
\[
\ovcalT^{(1, 0, 0)} (\zeta) = \calT^{(1, 0, 0)}(q^{3/s}\zeta), \qquad 
\ovcalT^{(1, 1, 0)}(\zeta) = \calT^{(1, 1, 0)}(q^{1/s}\zeta),
\]
following actually from the comparison of the representations $\varphi^\lambda_\zeta$ 
and $\ovvarphi^\lambda_\zeta$.

\section{Conclusions}

We have presented a short review of the universal functional $TQ$- and $TT$-relations for the quantum integrable systems 
associated with the quantum affine algebra $U_q(\calL(\gothsl_n))$, emphasizing the lower-rank ($n=2$) and a higher-rank 
($n=3$) cases as basic examples. We have also given the quantum analog of the Jacobi--Trudi identity allowing one to
express `higher-weight' universal transfer operators by means of two non-trivial basic operators with lowest weight. 
Here, the representation of the quantum group in the auxiliary space is properly specified, giving rise to this or 
another universal integrability object, while the representation in the quantum space is not. 

To consider concrete physical models, one should make the corresponding specialization of the quantum group also in the 
quantum space. Thus, choosing an appropriate infinite- or finite-dimensional representation of the quantum group in the 
quantum space, one can consider either a low-dimensional quantum field theory or a lattice model with the corresponding 
spin-chain counterpart. Upon such a specialization, one can use the remarkable Khoroshkin--Tolstoy formula \cite{KhoTol92, 
TolKho92, KhoTol93, KhoStoTol95} for the universal $R$-matrix to carry out explicit calculations of the monodromy operators 
\cite{Raz13, NirRaz14}, $R$-operators \cite{LevSoiStu93, ZhaGou94, BraGouZhaDel94, BraGouZha95, BooGoeKluNirRaz10, 
BooGoeKluNirRaz11, NirRaz14} and $L$-operators \cite{BooGoeKluNirRaz10, BooGoeKluNirRaz11}. Another possibility is
offered by a $3D$ approach to the Yang--Baxter equation \cite{Man14}.

\vskip .5em

{\em Acknowledgements.\/} This work was supported in part by the German Research Foundation grant KL \hbox{645/10-1}, 
the Russian Foundation for Basic Research grants \#~13-01-00217, \#~14-01-91335, and by the Volkswagen Foundation. 
We thank our colleagues and coauthors H. Boos, F. G\"ohmann and A. Kl\"umper for discussions. Kh.S.N. thanks the 
organizers of the ISQS'23 in Prague (June 23--27, 2015) for invitation and hospitality.

\appendix

\section{Defining $U_q(\gothgl_n)$ module relations} \label{a:1}

The root system of type $A_{n-1}$ consists of $n(n-1)$ roots. We introduce for the system
of positive roots the {\em normal ordering\/} \cite{AshSmiTol79, Tol89}
\begin{multline}
(\alpha_1), (\alpha_1 + \alpha_2, \, \alpha_2), (\alpha_1 + \alpha_2 + \alpha_3, \,
\alpha_2 + \alpha_3, \, \alpha_3), \ldots \\
\ldots (\alpha_1 + \alpha_2 + \ldots + \alpha_i, \, \alpha_2 + \ldots + \alpha_i, \ldots, \,
\alpha_i), \ldots \\
\ldots (\alpha_1 + \alpha_2 + \ldots + \alpha_{n-1}, \, \alpha_2 + \ldots + \alpha_{n-1}, \ldots, \, \alpha_{n-1}),
\end{multline}
where $\alpha_j$, $j = 1,\ldots,n-1$ are the simple positive roots. As usual, we define the root vectors corresponding to the negative composite roots as follows. Let $\gamma = \alpha + \beta$ be a composite positive root. Then the relation
\[
F_\gamma = F_\beta F_\alpha - q^{} F_\alpha F_\beta
\]
gives the root vector $F_\gamma$ corresponding to the negative root $-\gamma$. The whole 
set of negative root vectors ordered in accordance with the above given normal ordering
of the roots is used to define the basis vectors of the $U_q(\gothgl_n)$ modules. We write
\[ 
v_k = 
(F_1^{k_1}) (F_{12}^{k_{12}} F_2^{k_2}) (F_{13}^{k_{13}} F_{23}^{k_{23}} F_3^{k_3}) 
\cdots (F_{1i}^{k_{1i}} F_{2i}^{k_{2i}} \cdots F_i^{k_i}) \cdots 
(F_{1,n-1}^{k_{1,n-1}} F_{2,n-1}^{k_{2,n-1}} \cdots F_{n-1}^{k_{n-1}}) \, v_0,
\] 
where $k$ means the set of non-negative integers 
$(k_1$; $k_{12},k_2$; $k_{13},k_{23},k_3$; $\ldots$ $k_{1i},k_{2i},$ $\ldots,k_i$; 
$\ldots$ $k_{1,n-1},k_{2,n-1}$, $\ldots,k_{n-1})$ being the powers of the root vectors 
acting on the highest-weight vector $v_0$. Here we use the notation $F_{i}$ for the root 
vector corresponding to the simple negative root $-\alpha_i$, and $F_{ij}$ means the root
vector corresponding to the composite negative root $-\alpha_i - \ldots - \alpha_j$
which is defined according to the above relation.

Acting on the basis vectors $v_k$ by the generators $q^{\nu G_i}$, $E_i$ and $F_i$
of $U_q(\gothgl_n)$, we obtain the defining relations of the $U_q(\gothgl_n)$ modules.
For the simplest lower-rank case $n = 2$ the basis vectors are $v_k = F^k v_0$, and
we obtain
\begin{gather}
q^{\nu G_1} v_k = q^{\nu(\lambda_1 - k)} v_k, \qquad 
q^{\nu G_2} v_k = q^{\nu(\lambda_2 + k)} v_k, \label{a1}\\
F v_k = v_{k + 1}, \qquad E v_k = [k]_q [\lambda_1 - \lambda_2 - k + 1]_q v_{k-1}.
\label{a2}
\end{gather} 
For a more complicated and rich higher-rank case $n = 3$ we act on 
$v_k = F_1^{k_1} F_{12}^{k_{12}} F_2^{k_2} v_0$ and obtain
\begin{align}
& q^{\nu G_1} v_k = q^{\nu(\lambda_1 - k_1 - k_3)} v_k, \quad 
q^{\nu G_2} v_k = q^{\nu(\lambda_2 + k_1 - k_2)} v_k, \quad
q^{\nu G_3} v_k = q^{\nu(\lambda_3 + k_2 + k_3)} v_k, \label{a3}\\
& F_1 v_k = v_{k + \varepsilon_1}, \quad F_2 v_k =  q^{k_1 - k_3} v_{k + \varepsilon_3} + [k_1]_q v_{k - \varepsilon_1 + \varepsilon_2}, \quad F_3 v_k = q^{- k_1} v_{k + \varepsilon_2},
\label{a4}
\end{align}
\begin{align}
E_1 v_k & = [\lambda_1 - \lambda_2 - k_1 + k_2 - k_3 + 1]_q [k_1]_q v_{k - \varepsilon_1} -
 q^{\lambda_1 - \lambda_2  + k_2 - k_3 + 2} [k_2]_q v_{k - \varepsilon_2 + \varepsilon_3}, 
 \label{a5} \\
E_2 v_k & = [\lambda_2 - \lambda_3 - k_2 + 1]_q \, [k_2]_q v_{k - \varepsilon_3} + q^{- (\lambda_2 - \lambda_3) + 2 k_2} [k_3]_q 
v_{k + \varepsilon_1 - \varepsilon_2}, \label{a6} \\
E_3 v_k & = q^{k_1} [\lambda_1 - \lambda_3 - k_1 - k_2 - k_3 + 1]_q [k_3]_q v_{k - \varepsilon_2} \nonumber\\
& \hspace{3em} - q^{- (\lambda_1 - \lambda_2) + k_1 - k_2 + k_3 - 1} [\lambda_2 - \lambda_3 - k_2 + 1]_q [k_1]_q [k_2]_q 
v_{k - \varepsilon_1 - \varepsilon_3}. \label{a7}
\end{align}  
Here, for brevity, we have denoted $F_{12}$ and $k_{12}$ by $F_3$ and $k_3$, respectively,
and use the notation $k - \varepsilon_i$, $i = 1,2,3$, for the sets of three integers 
$(k_1-1,k_3,k_2$), $(k_1,k_3-1,k_2$) and $(k_1,k_3,k_2-1$), respectively.

Thorough consideration of these module relations and appropriate interpretation of
their consequences is the actual basis for the complete proof of the universal
functional relations.

\bibliographystyle{amsrusunsrt}

\newcommand{\noopsort}[1]{}
\providecommand{\bysame}{\leavevmode\hbox to3em{\hrulefill}\thinspace}
\providecommand{\href}[2]{#2}
\providecommand{\curlanguage}[1]{%
 \expandafter\ifx\csname #1\endcsname\relax
 \else\csname #1\endcsname\fi}


\end{document}